\documentclass[superscriptaddress,twocolumn]{revtex4}

\usepackage{amsfonts}
\usepackage{amsmath}    
\usepackage[normalem]{ulem}
\usepackage{bm}
\usepackage{graphicx}

\newcommand{\ket}[1]{\vert#1\rangle}

\def\opone{\leavevmode\hbox{\small1\kern-3.8pt\normalsize1}}
\usepackage{color}

\usepackage{amsfonts, amsmath, amssymb, bm, bbm, graphicx, epsfig, epstopdf}
\usepackage{dcolumn}
\usepackage{textcomp}
\usepackage{hyperref}
\usepackage{color}

\begin{document}
\title{Cross-phase modulation of a probe stored in a waveguide for non-destructive detection of photonic qubits
}

\author{N. Sinclair}
\affiliation{Institute for Quantum Science and Technology, and Department of Physics and Astronomy, University of Calgary, Calgary T2N 1N4, Alberta, Canada}
\author{K. Heshami}
\affiliation{National Research Council of Canada, 100 Sussex Drive, Ottawa, Ontario, K1A 0R6, Canada}
\author{C. Deshmukh}
\affiliation{Institute for Quantum Science and Technology, and Department of Physics and Astronomy, University of Calgary, Calgary T2N 1N4, Alberta, Canada}
\author{D. Oblak}
\affiliation{Institute for Quantum Science and Technology, and Department of Physics and Astronomy, University of Calgary, Calgary T2N 1N4, Alberta, Canada}
\author{C. Simon}
\affiliation{Institute for Quantum Science and Technology, and Department of Physics and Astronomy, University of Calgary, Calgary T2N 1N4, Alberta, Canada}
\author{W. Tittel}
\affiliation{Institute for Quantum Science and Technology, and Department of Physics and Astronomy, University of Calgary, Calgary T2N 1N4, Alberta, Canada}

\maketitle

{\bf Non-destructive detection of photonic qubits is an enabling technology for  quantum information processing and quantum communication \cite{Imoto,matsuda,bajcsy,shiau,lam,reiserer2013a,gaeta,Kalb,Steinberg,Franson}. For practical applications such as quantum repeaters \cite{Sangouard} and networks \cite{Gisin,Kimble}, it is desirable to implement such detection in a way that allows some form of multiplexing as well as easy integration with other components such as solid-state quantum memories \cite{Lvovsky,Bussieres}. Here we propose an approach to non-destructive photonic qubit detection that promises to have all the mentioned features. Mediated by an impurity-doped
crystal, a signal photon in an arbitrary time-bin qubit state \cite{Weihs} modulates the phase of an intense probe pulse that is stored during the interaction. Using a thulium-doped waveguide in LiNbO$_3$, we perform a proof-of-principle experiment with macroscopic signal pulses, demonstrating the expected cross-phase modulation as well as the ability to preserve the coherence between temporal modes. Our findings open the path to a new key component of quantum photonics based on rare-earth-ion doped crystals.}

The ability to detect photonic qubits non-destructively is highly desirable for photonic quantum information processing and quantum communication. For instance, it makes it possible to use precious resource states (say entangled photon pairs for quantum teleportation) only when the input photons are actually there. This is all the more essential in situations of significant loss, such as for quantum repeaters \cite{Sangouard,Boone}. Non-destructive detection of photons \cite{reiserer2013a} and heralded storage of photonic qubits \cite{Kalb} (which, when combined with readout, is equivalent to non-destructive detection) have recently been realized in sophisticated quantum electrodynamics experiments that combine single-atom control and high-finesse cavities, and work at one specific atomic transition frequency. For practical applications it is important to have a simple and robust implementation of the same functionality but with added flexibility. In particular, it should  allow for multiplexing, and be compatible with existing quantum information processing and communication components.

\begin{figure}[t]
\includegraphics[width=\columnwidth]{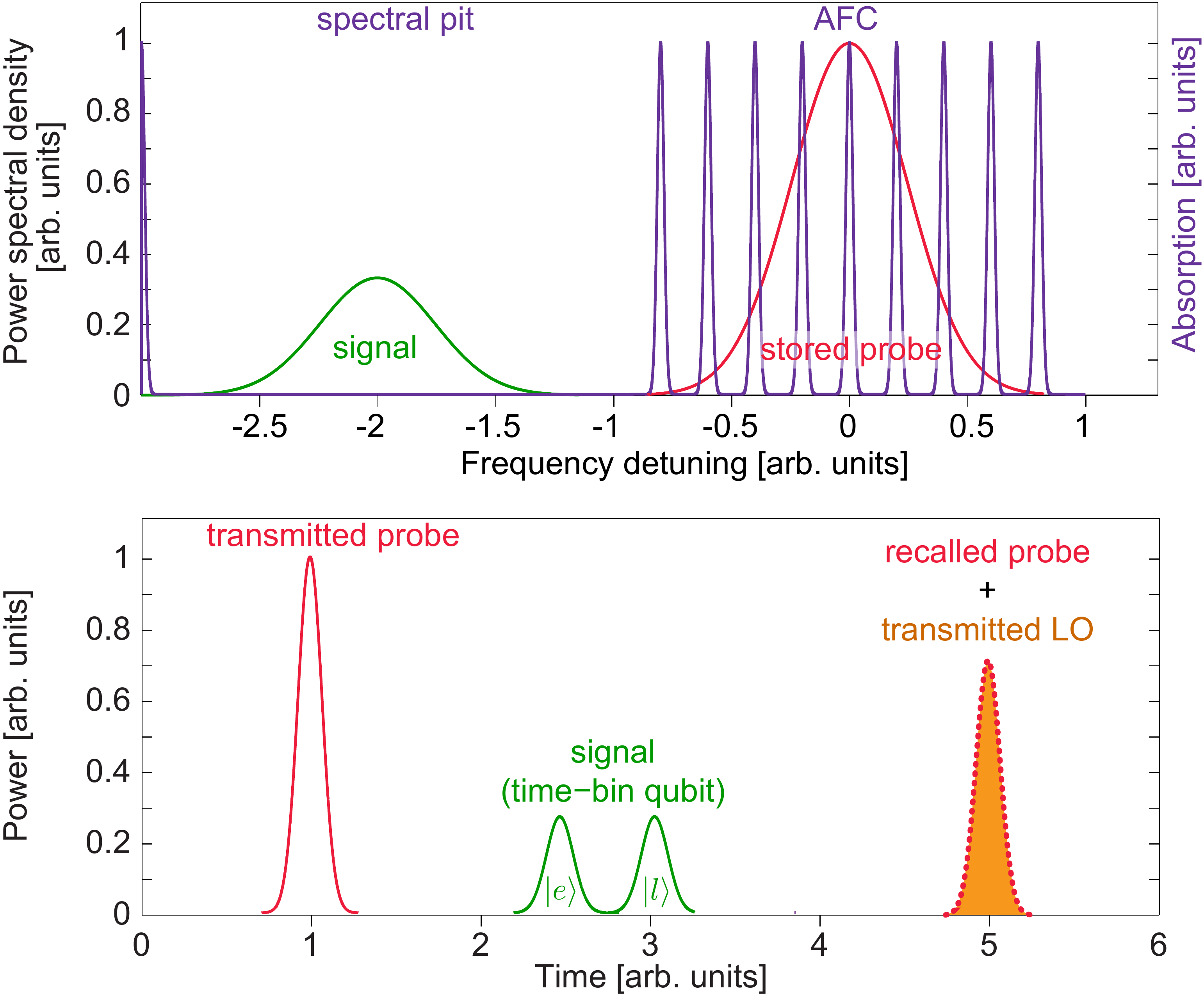}
\caption{\textbf{Non-destructive detection of photonic time-bin qubits.} A macroscopic probe pulse is stored in an atomic frequency comb (AFC) memory. The signal --- a photonic time-bin qubit --- propagates through a detuned transparency window and frequency shifts the atoms constituting the AFC due to the AC Stark effect. This results in a phase shift of the re-emitted probe. \textbf{a,} Spectral representation. \textbf{b,} Temporal representation. $\ket{e}$ and $\ket{l}$ denote early and late qubit modes, respectively. }\label{fig:scheme}
\end{figure}

\begin{figure*}[t]
\includegraphics[width=\textwidth]{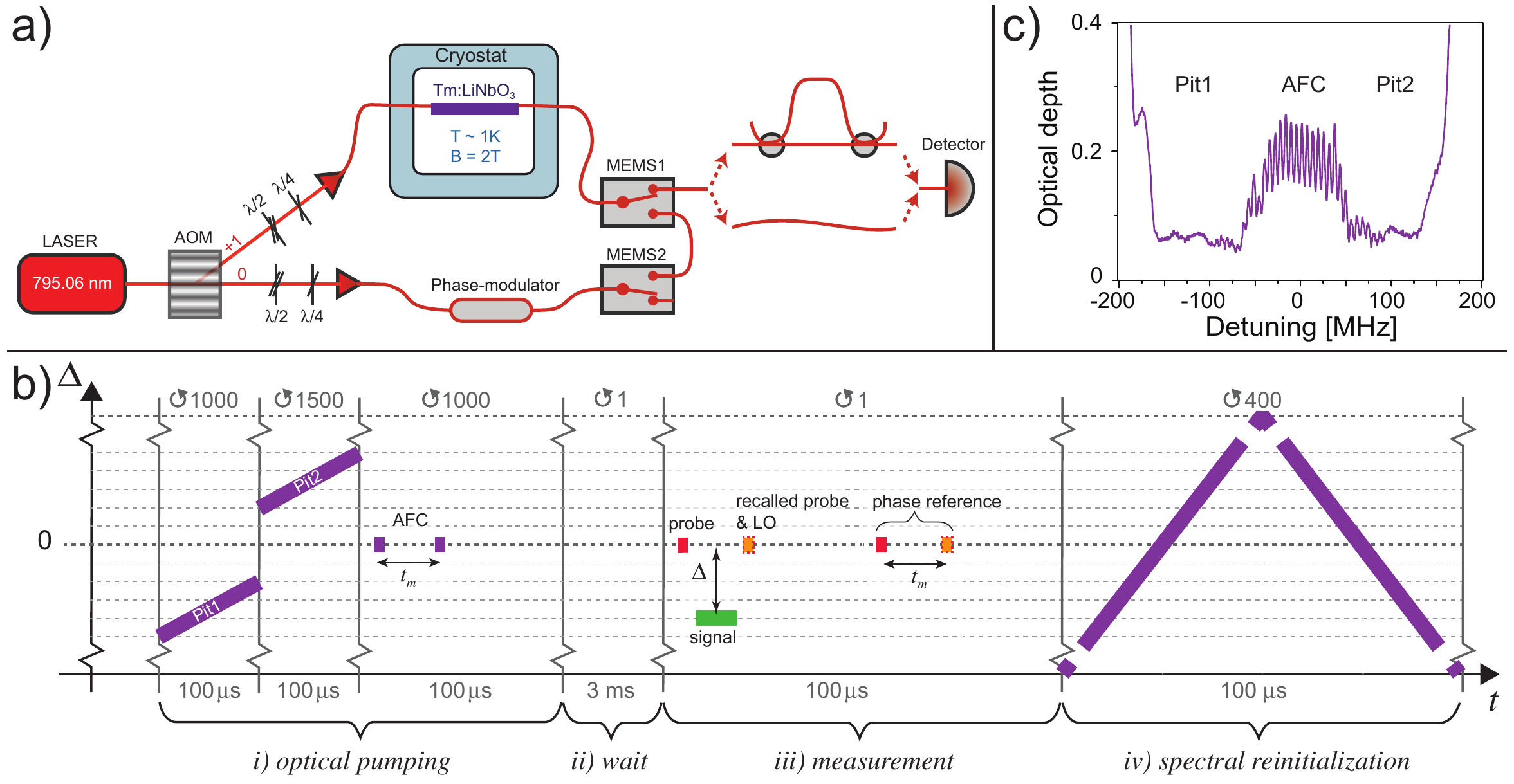}
\caption{\textbf{Outline of the experiment.}
\textbf{a,} Setup. Light from a frequency-locked 795.06~nm CW-laser is intensity and frequency modulated using an acousto-optic modulator (AOM). The diffracted first-order beam is coupled via fiber into the Tm:LiNbO$_3$ waveguide, and waveplates enable adjusting its polarization to maximize the interaction with Tm ions within a single spatial mode of $\sim$12.5~$\mu$m diameter (characterized independently). \textbf{b,} Timing sequence. \emph{i) Optical pumping} involves repetitive spectral pit burning at negative (-150 to -50~MHz) and positive (50 to 150~MHz) detunings for at total of 250~ms, and AFC generation using many pulse-pairs for 100~ms. (Depicted is one repetition as well as the number of repetitions per task.) \emph{ii)} After a 3~ms \emph{wait} time to allow excited atomic population to decay, we perform our \emph{measurement (iii)}: A 10~ns long probe is stored in the AFC, followed by a detuned signal that is transmitted through a spectral pit. A local oscillator (LO) interferes with the probe pulse recalled after 180~ns storage. Another 200~ns later, we perform a phase reference measurement using the same sequence but excluding the signal pulse.
At the memory output, a micro electro-mechanical switch (MEMS1) blocks light during optical pumping. It opens during the measurement to allow the transmission of the recalled probe pulse to the detector --- either directly or via an unbalanced interferometer, depending on the measurement performed.
\emph{iv)} As the strong probe pulses modify the tailored spectral feature, we \emph{reinitialize} the absorption line after every measurement using zeroth-order light from the AOM that is repetitively frequency-modulated over a 5-GHz range by a phase modulator. The light enters the thulium-doped waveguide through MEMS2 and MEMS1; it is blocked by MEMS2 outside the reinitialization step of 40 ms duration.
\textbf{c,} Spectral feature. A 100~MHz wide AFC with a tooth separation $\Delta_m/(2\pi)$ = 5.5~MHz (corresponding to a storage time of $t_m=180$~ns) and a 100~MHz wide spectral pit on either side of the AFC.}
\label{fig:setup}
\end{figure*}

Here we propose a detection scheme that has all of these characteristics. The basic principle, illustrated in Fig.~\ref{fig:scheme}, is based on cross-phase modulation between a weak signal and a strong probe pulse mediated by a rare-earth ion doped crystal --- a technology platform whose suitability for quantum photonics has already been demonstrated \cite{Bussieres,spin,Hedges,Clausen,Saglamyurek,cavityAFC,Sinclair,processor}. For single-photon sensitivity, the phase shift has to be greater than the quantum phase uncertainty of the probe, which is of order $1/\sqrt{N_p}$, where $N_p$ is the number of photons in the probe.
The probe is stored in an impurity-doped crystal using the atomic frequency comb (AFC) quantum memory protocol \cite{Afzelius}, and the phase shift is due to the AC Stark shift of the relevant atomic transition caused by the signal. For large detuning between signal and probe, it is given by
\begin{equation}
\phi=N_s\frac{1}{4\pi}\frac{\lambda^2 }{n^2 A }\frac{\gamma}{\Delta},
\label{eq:phase}
\end{equation}
where $N_s$ is the number of photons in the signal, $\lambda$  the vacuum wavelength of the atomic transition, $n$ the refractive index of the crystal, $A$ the interaction cross section, $\gamma$ the spontaneous decay rate from the excited state,  and $\Delta /(2\pi)$ the detuning in Hz.  See the Supplementary Information for a detailed derivation. Eq.~(\ref{eq:phase}) shows that the phase shift benefits from lateral confinement (small $A$) and small detuning, and that it increases linearly with the number of signal photons.

We emphasize that the phase shift does not depend on the exact timing of the signal, as long as it propagates through the medium while the probe is being stored. In particular, this allows one to detect the presence of a photon without affecting its qubit state, provided that the qubit is encoded in temporal modes --- a very convenient and widely-used choice in quantum communication. (Note that photonic qubits can easily be converted between different types of encoding \cite{UTBA}).

Our experimental set-up, sketched in Fig.~\ref{fig:setup}a, is composed of a Tm:LiNbO$_3$ waveguide quantum memory, a source for signal and probe pulses, and analyzers that allow characterizing these pulses after the waveguide-mediated interaction. We use the optical pumping sequence illustrated in Fig.~\ref{fig:setup}b to spectrally tailor the inhomogeneously broadened ${}^3H_6\rightarrow {}^3H_4$ absorption line of Tm into a series of absorption peaks (teeth) spaced by angular frequency $\Delta_m$ (the AFC), surrounded by transparent pits (see Fig~\ref{fig:setup}c). The bandwidth of the AFC and each of the pits is about 100~MHz, and the AFC memory storage time, given by $t_m=2\pi/\Delta_m$, is 180~ns.


Following the spectral tailoring, we generate a probe pulse of $\sim10$~ns duration whose spectrum matches the AFC. A part of the pulse is transmitted through the waveguide and a part of it is stored in the thulium ions forming the AFC. As illustrated in Fig~\ref{fig:setup}b, we then send a signal whose temporal structure, intensity and detuning w.r.t. the AFC we can vary, depending on the desired measurement. After the storage time $t_m$ the probe pulse is re-emitted from the memory. To measure its phase change due to the interaction with the signal, we interfere it with a local oscillator (LO). See the Methods section for more details about the AFC generation and the measurement.
\begin{figure}[t]
\includegraphics[width=\columnwidth]{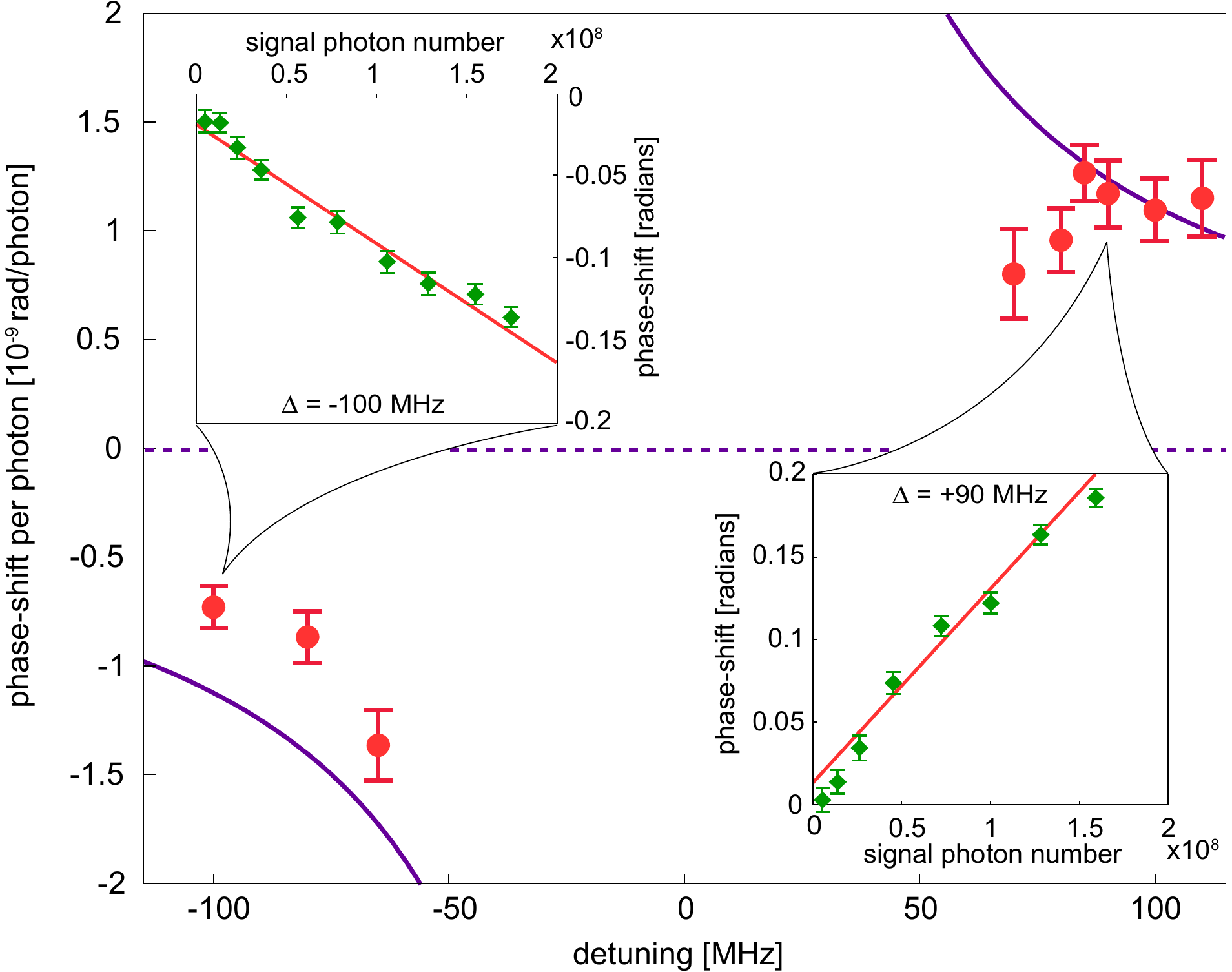}
\caption{\textbf{Phase shift per photon for different detuning values}. Expected phase shifts (purple line) according to Eq.~\ref{eq:phase} (no fit), and experimentally obtained values (red circles) derived from linear fits  to the phase shift vs. mean photon number as illustrated in the insets for two detuning values (red lines). Each data point in the insets (green diamonds) corresponds to an average over 200~repetitions. Uncertainty bars indicate the standard deviation of the average. Discrepancies between measured and predicted values are most likely due to imperfect AOM operation resulting in non-ideal signal spectra and pits (see Fig. \ref{fig:setup}c), both of which are asymmetric w.r.t. zero-detuning.
}
\label{fig:phase-vs-power-and-detune}
\end{figure}

First, to verify the probe-phase-shift dependence given in Eq.~\ref{eq:phase}, we use a signal pulse in a single temporal mode of 130 ns duration. We vary the number of photons per pulse for nine different detunings, and record the phase shift averaged over 200 repetitions for each photon-number.
As expected, we find a linear increase as a function of the number of signal photons, and that the slopes for red and blue detuning have opposite signs, as shown for two detunings in the inserts of Fig~\ref{fig:phase-vs-power-and-detune}. From the fitted slopes we find the phase-shifts per photon, which are shown in Fig~\ref{fig:phase-vs-power-and-detune} together with the expected values. We see that the measured data closely follows the theoretical predictions derived from Eq.~\ref{eq:phase} using $\lambda$=795 nm, $n$=2.3, A=$\pi\times$(6.25 $\mu$m$)^2$,  $\gamma$=9.1~kHz. In particular, at +100 MHz detuning, we measure a phase shift of $1.10\times10^{-9}$~rad/photon, which is in excellent agreement with the expected value of $1.12\times10^{-9}$~rad/photon.

\begin{figure}[t]
\includegraphics[width=\columnwidth]{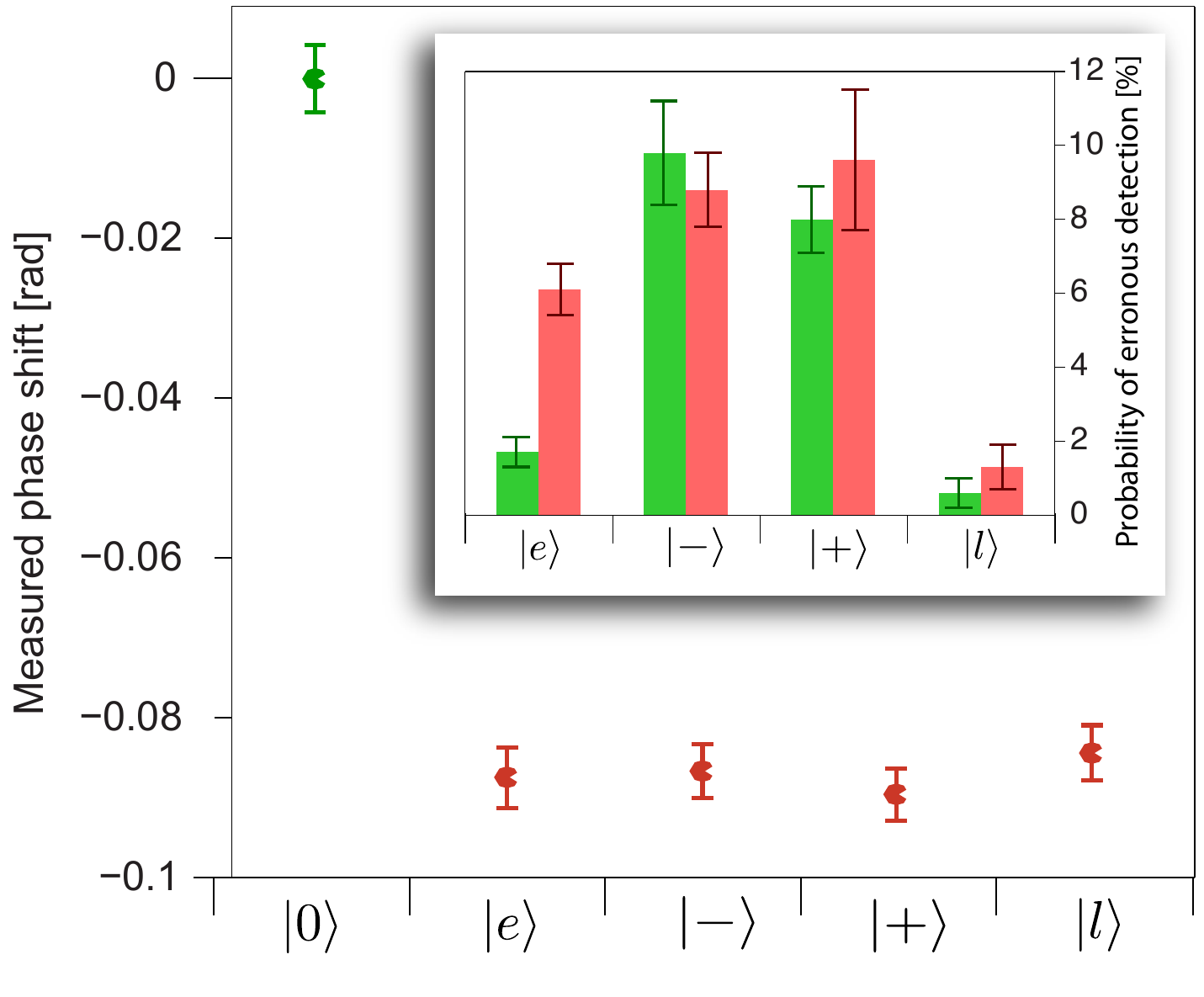}
\caption{\textbf{State preservation for signals in different temporal modes}. Probe phase shifts due to 6.9$\times$10$^7$, or no, signal photons, distributed between early and late temporal modes. The labels on the x-axis refer to corresponding time-bin qubit states. Each data point shows the average over 1000 measurements, and uncertainty bars denote the standard deviation of the average. The inset shows the error rates of the different signal states before and after the measurement (error bars are calculated from shot-to-shot pulse-heights variations). There is no significant change, except for $\ket{e}$. (Increased errors are likely due to free induction decay after exciting remaining thulium atoms inside the pit, and would disappear with better hole burning. As the decay happens after absorption, only $\ket{e}$ is affected. Errors for the superposition states are caused by imperfections in the interferometer.)
}
\label{fig:phase-qubit}
\end{figure}

Next, we demonstrate that the probe phase shift does not depend on how the signal energy is distributed between two temporal modes, and that the signal is not affected by the measurement. Put into the context of an interaction with a single photon in a time-bin qubit state, this implies that the measurement does not project the qubit onto a specific set of basis states and thus alter it. Towards this end, we select early and late signal modes, each of 10~ns duration, separated by 18.3~ns, and featuring a detuning of +100~MHz. Keeping the total energy constant, we generate signals in which the energy is concentrated in either the early or the late mode, or in an equal superposition with either 0 or $\pi$ phase-difference ('+' and '-' superpositions, respectively). The resulting probe phase shifts, averaged for each pulse sequence over 1000 repetitions, are plotted in Fig.~\ref{fig:phase-qubit}, which also includes the phase shift measured without a signal pulse. We find that, within experimental uncertainty, the phase shifts are the same irrespective of the signal state, and they clearly differ from the phase shift measured without any signal. Furthermore, to verify that our measurement preserves the signal state, we assess erroneous detections of signals prepared in various states without and with the measurement (see the Methods section for details). As shown in the inset of Fig.~\ref{fig:phase-qubit}, we find close to no change due to the cross-phase interaction, which is consistent with the fact that our scheme can measure the presence of a time-bin qubit without revealing, nor modifying, its state.

While our proof-of-principle demonstration confirms the key features of the proposed  scheme, a lot remains to be done before qubits encoded into individual and spectrally multiplexed photons can be detected non-destructively and without averaging. We expect that a reduction of the interaction cross section, e.g. using a small-diameter ridge waveguide, can improve the phase sensitivity by more than a factor of 100 (see Eq.~\ref{eq:phase}). Furthermore, the ratio between the radiative lifetime $\gamma$ and the detuning $\Delta$ has to be increased beyond its current value of 9~kHz/(2$\pi\times$65 MHz)$\sim$2.2$\times$10$^{-5}$ --- it can in principle approach a percent. 
As a result of these improvements, the phase shift per photon could thus be as large as 100 $\mu$rad, which would allow single-shot detection of individual photons.

Reducing the detuning to maximize $\gamma /\Delta$ comes with the unwanted effects of increasing off-resonant absorption of the signal in the AFC, increasing the noise due to decay from excited atoms, and decreasing the achievable bandwidth. However, as we discuss in more detail in the Supplementary Information, these problems can be overcome in a configuration in which the population in the excited state (populated through the absorption of the strong probe in the AFC) is temporarily transferred to an auxiliary level, and in which the signal passes many times through the spectral pit during the storage of the probe (using, e.g., a cavity \cite{cavityAFC}). This makes it possible to increase the detuning and thus reduce the absorption of the signal without decreasing the number of atoms in the AFC nor the total phase shift experienced by the probe. For instance, we anticipate the nondestructive measurement to be feasible for photons of half a MHz bandwidth, using an AFC with teeth of optical depth 30 \cite{Hedges}, and interacting approximately 900 times with the stored probe, which corresponds to a moderate-finesse cavity.

We emphasize that the cross-phase interaction in rare-earth-ion doped crystals is straightforward to generalize to multiple spectral channels, as demonstrated in the context of AFC-based optical quantum memory \cite{Sinclair}, which can extend over a total bandwidth of hundreds of GHz \cite{Charles}. We also note that the present approach should  allow the development of a standard (destructive) photon-number-resolving detector, for which the limitations imposed by signal loss and noise are less severe.

We believe that an improved version of our proof-of-principle demonstration will soon allow first destructive, and then non-destructive, ``single shot" detection of photons. This will open the path to more efficient use of precious resources in advanced applications of quantum communication, and the heralded generation of photon-number states, including entangled states, that do not contain often detrimental admixtures of undesired photons as, e.g., in widely-used spontaneous parametric down-conversion \cite{Guha}.
Finally, we note that the bandwidth of the signal in our scheme is proportional to the linewidth of the relevant atomic transition, which, in our case, is around 10~kHz. To allow the non-destructive detection of photons featuring more than of order 1 MHz bandwidth, it may therefore be interesting to investigate impurity-doped crystals with transitions that feature shorter lifetimes, e.g. Ce:YAG \cite{Kolesov}.

\section*{Methods}

\noindent
\textbf{Spectral tailoring.}
We tailor the spectrum of the inhomogeneously broadened ${}^{3}H_{6} \rightarrow {}^{3}H_{4}$ absorption line in Tm$^{+3}$ by means of frequency-selective optical pumping of  Tm-ions to another ground-state Zeeman level \cite{Charles}. The Zeeman splitting in the applied 2~T field is $\sim$2.5~GHz, which sets the upper limit for the total width of our spectral feature (the AFC and two pits). However, the bandwidth of the AOM used for laser frequency modulation practically limits the total width to 300~MHz.

\noindent
First, as illustrated in Fig.~\ref{fig:setup}b, we generate two spectral pits by sweeping the frequency of the laser light repeatedly over two 100~MHz wide regions separated by a spectral interval of 100 MHz. The optical depth of the remaining background is around 0.07~dB. It is irregular due to varying efficiency of the AOM with detuning.
Next, we generate an AFC in between the two pits by driving the AOM using pairs of pulses that are 10~ns wide and separated by 180~ns. The resulting AFC features a bandwidth of 100 MHz and a tooth separation of $\Delta_m/(2\pi) = 5.5$~MHz, corresponding to a storage time of 180~ns. The tooth separation is chosen to match side-peaks at 11~MHz 
arising from the super-hyperfine interaction of thulium with niobium in the host crystal [NS \textit{et al.}, in preparation], and the teeth width is limited by the long-term laser linewidth (1~MHz) combined with power broadening during spectral tailoring. The teeth feature an optical depth of $\sim$0.1, and are sitting on a background with optical depth of $\sim$0.15, resulting in a recall efficiency for the probe of 0.2\% \cite{Afzelius}.

\noindent
The quality of our spectral feature --- the background in the pits and the AFC, as well as the small optical depth of the AFC teeth --- is currently limited by long-term laser frequency jitter and non-ideal spectral tailoring. It can be improved by using a laser with improved stability, and by optical pumping based on ``burning back population from the side" \cite{Hedges}. This will allow meeting the requirements for a non-destructive measurement at the single photon level detailed in the Supplementary Information.\\

\noindent
\textbf{Measurements.}
\emph{Phase measurements --} Assessing the cross-phase modulation relies on an interferometric measurement of the recalled probe pulse with a transmitted local oscillator (LO) in  the same spatial, temporal and spectral mode, and featuring the same intensity. First, by varying the phase of the LO in the absence of a signal, we calibrate the interference visibility to 89.7\%. Next, to ensure maximum measurement sensitivity, we set the phase difference between the LO and the recalled probe (still without a signal) to $\pi/2$. Taking the calibration into account, this allows us in the actual measurement to map intensities (after interfering the probe with the LO) onto phase changes of the probe. Please note that the intensity of the recalled probe does not depend on whether or not a signal is present, i.e. the calibration, taken without any signal, remains valid when the latter is present.

\noindent
The precision of the phase measurements is mainly limited by long-term laser frequency instability.
We estimate that fluctuations between the AFC generation and the creation of the probe $\sim$3 ms later result in shot-to-shot noise of around 150~mrad.
To reduce this noise, we concatenate each measurement of the AC Stark shift on the probe with a reference measurement of the probe's phase without a signal (see Fig. \ref{fig:setup}b). Subtracting the values obtained by these two phase-measurements (with a weight given by the correlation of two subsequent measurements without signal) allows improving the single-shot phase sensitivity to around 100~mrad. This value is mainly limited by laser frequency fluctuations between the generation of the probe and LO pulses, and can be further reduced by improved laser locking. In addition, pulse intensity fluctuations and electronic noise of the photo-detector contribute $\sim$50~mrad of phase uncertainty.
By averaging phases over $j$ measurement repetitions, the sensitivity improves by a factor of $j^{-1/2}$. For instance, for $j=200$, we reach a  resolution of $\sim$7~mrad. 

\noindent
\emph{Qubit measurements --}
The variation of the signal due to the interaction with the probe is assessed as follows: for early and late signal states we measure the pulse heights in the wrong time bin, normalized to the sum of the pulse heights in both bins. For the superposition states, we pass the signal through an imbalanced fiber interferometer whose arm-length difference corresponds to 18.3 ns travel time difference. Using a piezoelectric transducer in one arm of the interferometer, we set its phase to obtain maximum constructive interference in one output, and record the normalized pulse heights in the other (the wrong) output. All measurements are done twice --- once before, and once after the signal is submitted to the cross-phase interaction. Differences in the results indicate the perturbation of the signal due to the measurement.

 \vspace{1cm}

\noindent
\textbf{Acknowledgements}\\
The authors thank Erhan Saglamyurek and Mikhail Lukin for useful discussions, and acknowledge funding through Alberta Innovates Technology Futures (AITF), the National Science and Engineering Research Council of Canada (NSERC), and the Defense Advanced Research Projects Agency (DARPA) Quiness program (contract no. W31P4Q-13-l-0004). W.T. acknowledges funding as a Senior Fellow of the Canadian Institute for Advanced Research (CIFAR).  \\

 \noindent
\textbf{Additional information}\\
Supplementary information is available in the online version of the paper. Correspondence and requests for materials should be addressed to W. Tittel (email: \mbox{wtittel@ucalgary.ca)} and/or C. Simon  (email: \mbox{csimo@ucalgary.ca}).\\

\noindent
\textbf{Competing financial interests}\\
The authors declare no competing financial interests.
\\

\pagebreak
\widetext

\begin{center}
\textbf{\large SUPPLEMENTARY INFORMATION}
\end{center}
\renewcommand{\theequation}{SI\arabic{equation}}

\setcounter{table}{0}
\setcounter{equation}{0}
\setcounter{figure}{0}
\makeatletter
\renewcommand{\bibnumfmt}[1]{[SI#1]}
\renewcommand{\citenumfont}[1]{SI#1}
\renewcommand{\figurename}[1]{FIG. SI#1}

\section{Storage of the probe; use of additional levels}
\label{probestorage}

The first step in our protocol involves the storage of a classical probe field. Here, we provide a semi-classical treatment of the light-matter interaction to describe AFC storage and retrieval of the probe field (see also \cite{Afzelius2009}). The total Hamiltonian describing our system is given by

\begin{equation}\label{Htot}
{\hat H}={\hat H}_0+{\hat H}_{int},
\end{equation}
where
\begin{equation}\label{H0}
{\hat H}_0=\sum_{j=1}^{N}{\hbar \omega_{ge}^{j}{\hat \sigma}_{ee}^{j}},
\end{equation}
and
\begin{equation}\label{Hint}
{\hat H}_{int}=-\hbar\sum_{j=1}^{N}{\left(\Omega (z,t) {\hat \sigma}_{eg}^{j} e^{-i\omega_p(t-z_j/c)}+ H.c.\right) }.
\end{equation}
Here, $\hbar \omega_{ge}^j$ denotes the excited state energy of atom $j$ and $\omega_p$ is the control frequency of the probe field; $\Omega(z,t)=\frac{\mu_{eg} {\mathcal E}_p(z,t)}{2\hbar}$ is the Rabi frequency associated with the probe field (${\mathcal E}_p$), and $\mu_{eg}=\langle e|{\hat d}.{\mathbf \epsilon_p}|g\rangle$ is the transition dipole moment.
Using the Heisenberg equation, one can find the following dynamical equations,
\begin{equation}\label{dyn1}
{\dot{\hat\sigma}}_{gg}(z,t;\delta)=i\Omega^*(z,t){\hat\sigma}_{ge}(z,t;\delta)-i\Omega(z,t){\hat\sigma}_{eg}(z,t;\delta),
\end{equation}
\begin{equation}\label{dyn2}
{\dot{\hat \sigma}}_{ee}(z,t;\delta)=i\Omega(z,t){\hat \sigma}_{eg}(z,t;\delta)-i\Omega^*(z,t){\hat\sigma}_{ge}(z,t;\delta),
\end{equation}
and
\begin{eqnarray}\label{dyn3}
{\dot{\hat\sigma}}_{eg}(z,t;\delta)=i(\omega_0+\delta-\omega_p){\hat\sigma}_{eg}(z,t;\delta)+i\Omega^*(z,t){\hat \sigma}_{ee}(z,t;\delta)
-i\Omega^*(z,t){\hat \sigma}_{gg}(z,t;\delta),
\end{eqnarray}
where
\begin{equation}\label{sig_gg}
{\hat\sigma}_{gg}(z,t;\delta)=\frac{1}{N_z(\delta)}\sum_{i=1}^{N_z(\delta)} {\hat\sigma}_{gg}^{i}(t;\delta),
\end{equation}
\begin{equation}\label{sig_ee}
{\hat\sigma}_{ee}(z,t;\delta)=\frac{1}{N_z(\delta)}\sum_{i=1}^{N_z(\delta)} {\hat\sigma}_{ee}^{i}(t;\delta),
\end{equation}
and
\begin{equation}\label{sig_eg}
{\hat\sigma}_{eg}(z,t;\delta)=\frac{1}{N_z(\delta)}\sum_{i=1}^{N_z(\delta)} {\hat\sigma}_{eg}^{i}(t;\delta)e^{-i\omega_p(t-z_i/c)},
\end{equation}
Here, $\omega_0$ is the central frequency of the inhomogeneously broadened atomic ensemble, $\omega_p$ is the central frequency of the probe field, and $\delta$ denotes the detuning of different modes of the ensemble with respect to its central frequency. The atomic coherence and population for atom $j$ are determined by ${\hat \sigma}_{\nu\nu'}=|\nu\rangle^j\langle\nu'|$, where $\nu,\nu'=g,e$; ${\hat\sigma}_{\nu\nu'}(z,t;\delta)$ are collective atomic operators for all atoms in a slice of the medium for (longitudinal) position $z$ and frequency mode $\delta$. Note that we assume that the number of atoms in mode $\delta$ at $z$, $N_z(\delta)$, is much larger than 1. In addition $N_(\delta)$ characterizes a periodic absorption feature (in the frequency domain) with periodicity $\Delta_m$ that is required for AFC storage. The propagation of the probe field can be derived starting from Maxwell's equations,
\begin{equation}\label{Max}
\left( \partial_z + \frac{n}{c}\partial_t\right){\mathcal E}_p(z,t)=\frac{i\mu_0\omega_p^2}{2k_p}\langle{\hat{\mathcal P}_{tot}}\rangle,
\end{equation}
where ${\mathcal E}_p(z,t)$ is the slowly varying envelope of the probe field, $k_p=\frac{n\omega_p}{c}$, and $\langle{\hat{\mathcal P}_{tot}}\rangle$ denotes the expectation value of
\begin{equation}\label{Ptot}
{\hat{\mathcal P}_{tot}}=\sum_{\delta} \langle g|{\hat d}.{\mathbf \epsilon_p}|e\rangle\frac{N(\delta)}{V}{\hat \sigma}_{ge}(z,t;\delta).
\end{equation}
Equations \ref{dyn1},\ref{dyn2},\ref{dyn3} along with Eqs. \ref{Max},\ref{Ptot} allow us to describe the dynamics of the atoms due to the probe field, when $t<T_1$ and $t>T_2$. $T_1<t<T_2$ is the time between the probe storage and retrieval in which the evolution in perturbed by presence of the signal field; see below. Storage of a probe field carrying an average photon number $N_p$ that is smaller than the total number of atoms $N_g$ is expected to result in a coherent state distribution of atomic excitations. In addition, the bosonic characteristics of the collective atomic excitation (${\hat \sigma}_{ge}(z,t;\delta)$) can be used to evaluate the above expectation value of the total atomic polarization operator.

In the proof-of-principle experiment reported in the main text the same atomic transition is used for signal and probe fields. To minimize loss and noise for the signal, it is desirable not to have significant population in the excited state $e$ when the signal propagates through the medium. This can be achieved either by using transitions from the same ground state to two different excited states (a V configuration), or by transferring the excited state population to another level (e.g. another ground state level or a metastable state, i.e. a $\Lambda$-type configuration) after the probe has been absorbed. The latter approach also provides a larger time window for signal propagation. Under these conditions, the signal sees an $e-g$ transition where there is no population in $e$, but the number of atoms in $g$ is reduced with respect to the total initial number by the number of probe photons that were absorbed. For optimum phase sensitivity the number of absorbed photons $N_p$ should be of order $N/2$, where $N$ is the total number of atoms. In this case the number of atoms remaining in the ground state $N_g$ is equal to $N_p$, $N_g=N_p=N/2$.

\section{Derivation of effective Hamiltonian for cross-phase modulation}

In this section, we provide the quantum-mechanical Hamiltonian for the interaction between the signal field and the atoms. For large detunings, where $\Delta$ is larger than the signal bandwidth, we derive an effective interaction Hamiltonian that will be used to find the probe phase shift with respect to the number of photons in the signal field.

The total Hamiltonian that governs the dynamics due to the presence of the signal field is given by
\begin{equation}\label{Hint_sig}
{\hat H}_{tot}={\hat H}_{0}+{\hat H}_{int}=\sum_{j=1}^{N} \hbar\omega_{ge}^j\sigma_{ee}^j+ {\hat h}_{int}^{j},
\end{equation}
where
\begin{equation}\label{hint_j}
{\hat h}_{int}^{j}=-\hbar g\sqrt{\frac{L}{2\pi c}}\int{d\omega {\hat a}_{\omega}e^{i\omega z_j/c}{\hat \sigma}_{eg}^j +\text{H.c.} }.
\end{equation}
The transition frequency of the $j^{\text{th}}$ atom is $\omega_{eg}^j$, and the atomic coherence and population operators are denoted by ${\hat \sigma}_{\nu\nu'}=|\nu\rangle^j\langle\nu'|$, where $\nu,\nu'=\{g,e\}$. The single photon coupling is given by $g=\mu_{eg}\sqrt{\frac{\omega_s}{2\hbar\epsilon V}}$, where $\omega_s$ is the central frequency of the signal and the transition dipole moment $\mu_{eg}= \langle e| {\bf d}\cdot {\bf \epsilon}_s|g\rangle$. We define ${\hat\sigma}_{\nu\nu'}(z,t;\delta)$ as collective atomic operators for all atoms in a slice of the medium at position $z$ and for frequency mode $\delta$. Here, $\omega_s$ is the central frequency of the signal field. Throughout this analysis, we assume that $\omega_p=\omega_0$, where $\omega_0$ is the central frequency of the AFC. This means that $\delta$ denotes different modes of the ensemble with respect to its central frequency.

We use the collective atomic operators that are defined in Eqs. \ref{sig_gg}, \ref{sig_ee} and \ref{sig_eg} to re-write the interaction Hamitonian in Eq. \ref{hint_j}. This results in
\begin{eqnarray}
{\hat H}_{int}= -\hbar g \sum_{j=1}^{N} \sqrt{\frac{L}{2\pi c}} e^{i\omega_p(t-z_j/c)} \int{d\omega {\hat a}_{\omega}e^{i\omega z_j/c}}{\hat \sigma}_{eg}^j e^{-i\omega_p(t-z_j/c)} +H.c.,
\end{eqnarray}
which leads to
\begin{eqnarray}\label{Hint_s}
{\hat H}_{int}= -\hbar g  \int dz n_z(\delta) e^{i\Delta(t-z/c)}{\hat {\mathcal E}}_s(z,t){\hat \sigma}_{eg}(z,t;\delta) +H.c.,
\end{eqnarray}
where ${\hat {\mathcal E}}_s(z,t)=\sqrt{\frac{L}{2\pi c}}e^{i\omega_s(t-z/c)}\int{d\omega {\hat a}_{\omega}e^{i\omega z/c}}$, $n_z(\delta)dz=N_z(\delta)$, $\int dz n_z(\delta)=N(\delta)$ and $\Delta=\omega_p-\omega_s$ is the detuning between the signal and probe fields.

For detunings much larger than the bandwidth of the signal field ($\Delta \gg 1/\tau_s$), we expect the dynamics of the atomic polarization (${\hat \sigma}_{eg}(z,t;\delta)$) to be dominated by the fast rotating terms of $e^{\pm i\Delta t}$. In order to capture this effect, let us consider the dynamics of ${\hat \sigma}_{eg}(z,t;\delta)$ due to ${\hat H}_{int}$. Starting from ${\dot {\hat \sigma}}_{eg}(z,t;\delta)=\frac{i}{\hbar}\left[{\hat H_{int}}, {\hat \sigma}_{eg}(z,t;\delta)\right]$, we find
\begin{eqnarray}
{\dot {\hat \sigma}}_{eg}(z,t;\delta)=-i g e^{-i\Delta (t-z/c)}{\hat {\mathcal E}}^{\dagger}_s(z,t)\left({\hat \sigma}_{gg}(z,t;\delta)-{\hat \sigma}_{ee}(z,t;\delta) \right),
\end{eqnarray}
which leads to
\begin{eqnarray}
{\hat \sigma}_{eg}(z,t;\delta)=-i g \int_{0}^{t} dt' e^{-i\Delta (t'-z/c)}{\hat {\mathcal E}}^{\dagger}_s(z,t')\left({\hat \sigma}_{gg}(z,t';\delta)-{\hat \sigma}_{ee}(z,t';\delta) \right).
\end{eqnarray}
For any state and for large detunings ($\Delta\gg 1/\tau_s$), this integral can be approximately evaluated by integrating the fast oscillating part and multiplying it by the final value of the slowly varying component. This approximation allows us to find the collective atomic polarization as
\begin{eqnarray}\label{sig}
{\hat \sigma}_{eg}(z,t;\delta)=\frac{g}{\Delta} e^{-i\Delta (t-z/c)}{\hat {\mathcal E}}^{\dagger}_s(z,t) \left({\hat \sigma}_{gg}(z,t;\delta)-{\hat \sigma}_{ee}(z,t;\delta) \right).
\end{eqnarray}
Using the above equation, in an iteration, we replace ${\hat \sigma}_{eg}(z,t;\delta)$ in Eq.\ref{Hint_s} to find an effective interaction Hamiltonian as follows,
\begin{eqnarray}\label{Hint_s_eff}
{\hat H}_{int}^{eff}= -\frac{\hbar g^2}{\Delta} \int dz n_z(\delta) \left({\hat {\mathcal E}}_s(z,t){\hat {\mathcal E}}^{\dagger}_s(z,t)+ {\hat {\mathcal E}}^{\dagger}_s(z,t){\hat {\mathcal E}}_s(z,t)\right) \left( {\hat \sigma}_{gg}(z,t;\delta)-{\hat \sigma}_{ee}(z,t;\delta)  \right).
\end{eqnarray}
Using the effective interaction Hamiltonian we derive the dynamical equation for the collective atomic polarization.
\begin{eqnarray}\nonumber
{\dot {\hat \sigma}}_{eg}(z,t;\delta)=\frac{i}{\hbar}\left[{\hat H_0}+{\hat H_{int}^{eff}}, {\hat \sigma}_{eg}(z,t;\delta)\right]+\frac{\partial {\hat \sigma}_{eg}(z,t;\delta)}{\partial t}.
\end{eqnarray}
This leads to
\begin{eqnarray}
{\dot {\hat \sigma}}_{eg}(z,t;\delta)=i\delta {\hat \sigma}_{eg}(z,t;\delta)+\frac{2ig^2}{\Delta} \left({\hat {\mathcal E}}_s(z,t){\hat {\mathcal E}}^{\dagger}_s(z,t)+ H.c.\right)  {\hat \sigma}_{eg}(z,t;\delta),
\end{eqnarray}
and consequently
\begin{eqnarray}\label{sig_T2}
& {\hat \sigma}_{eg}(z,t=T_2;\delta)=e^{i\delta t}e^{i{\hat \Phi}}{\hat \sigma}_{eg}(z,t=T_1;\delta),
\end{eqnarray}
where
\begin{eqnarray}\label{phi}
& {\hat \Phi}=\int_{T_1}^{T_2} dt' \frac{2g^2}{\Delta} \left({\hat {\mathcal E}}_s(z,t'){\hat {\mathcal E}}^{\dagger}_s(z,t')+{\hat {\mathcal E}}^{\dagger}_s(z,t'){\hat {\mathcal E}}_s(z,t')\right).
\end{eqnarray}
Note that the above dynamics describe the effects during the signal field propagation. The storage and retrieval of the probe field can be treated separately. Phase modulations due to the presence of the signal will appear in the first echo of the probe field. In addition, the total phase only depends on the total energy in the signal field and does not reveal any information about the temporal distribution of the signal field.

\section{ Phase shift per signal photon}

Given Eqs. \ref{sig_T2} and \ref{phi}, one can find the amount of phase shift for a single photon signal propagating in the waveguide and interacting off-resonantly with the atomic polarization. The electric dipole interaction Hamiltonian can be used to relate the spontaneous emission rate of a two-level system to its transition dipole moment. For two-level atoms in a solid with dipoles oriented along a specific direction, this results in $\gamma=\frac{\mu_{eg}^2 \omega_0^3}{\pi \epsilon \hbar c^3}$. Assuming that $\lambda_s\approx\lambda_0$ we can find the phase shift due to a single photon as
\begin{equation}\label{phi}
\phi=\frac{2g^2}{\Delta}\tau_s=\frac{1}{4\pi}\frac{\lambda_0^2}{n^2 A}\frac{\gamma}{\Delta},
\end{equation}
where $\tau_s=c/L$ is the duration of the signal in vacuum and $A$ is the cross-section area of the interaction, and $\lambda_0$ is the wavelength associated with the corresponding atomic transition in vacuum. If the population in the excited state is transferred to another ground state in order to minimize loss and noise (as discussed at the end of section I), then this expression for the phase shift has to be divided by a factor of 2 because this state is unaffected by the AC Stark shift due to the signal.

\section{Signal loss}

In this section, we analyze the signal photon loss due to its off-resonant interaction with the atoms in the AFC. In order to find a simplified description for off-resonant absorption loss, we assume that the signal detuning is larger than the inhomogenous bandwidth of the ensemble. This also guarantees that $\Delta$ is much larger than the spontaneous emission rate, $\gamma$.

For analyzing the signal loss, we treat the signal propagation using the Maxwell equation as follows
\begin{eqnarray}\label{MB_eqs}
\left( \partial_z+\frac{n}{c}\partial_t\right){\hat {\mathcal E}_s}(z,t)=\frac{\mu_0\omega_0^2\mu_{eg}}{2k_0}\sum_{\delta}{\frac{N(\delta)}{V} {\hat \sigma}_{eg}(z,t;\delta)},
\end{eqnarray}
where $k_0=\frac{n\omega_0}{c}$.
Given that the equations governing the single-excitation wave functions are the same as the Maxwell-Bloch equations, for evaluating the signal loss, we replace the atomic polarization and photonic annihilation operators with their corresponding wave functions. We are interested in the output signal wave function. This can be found by taking the Fourier transform of the Maxwell-Bloch equations. As a result the output signal can be written as
\begin{equation}\label{EM_FT}
{\tilde {\mathcal E}_s}(z,\omega)|_{z=L}=e^{ik_s\chi(\omega)L}{\tilde {\mathcal E}_s}(z=0,\omega),
\end{equation}
where
\begin{equation}\label{chi}
\chi(\omega)=\frac{1}{k_s}\left( -\frac{n\omega}{c}+\frac{\mu_0\omega_0^2\mu_{eg}}{2k_0}\sum_{\delta}{\frac{N(\delta)}{V}\frac{i\mu_{eg}/2\hbar}{i(\omega-(\Delta+\delta))-\gamma}}\right),
\end{equation}
and $k_s=\frac{n\omega_s}{c}$.
The imaginary part of $\chi(\omega)$ determines the loss. We can simplify the above expression by assuming $\Delta\gg\Gamma$, where $-\Gamma/2<\delta<\Gamma/2$. In addition, the loss is expected to be uniform over the signal field spectrum when its bandwidth is smaller than the detuning ($\Delta>1/\tau_s$). This results in a rather simple expression for the imaginary part of the response function. For $\Delta\gg\gamma$ this is given by
\begin{equation}\label{imag_chi}
\text{Imag}(\chi(0))=\frac{1}{k_s}\frac{1}{16\pi}\frac{N_g \lambda_0^2\gamma^2}{n^2 V\Delta^2}.
\end{equation}
Therefore, the intensity loss for the signal field can be derived from
\begin{equation}
|{\mathcal E}(L,\omega)|^2=e^{-\zeta L}|{\mathcal E}(0,\omega)|^2,
\end{equation}
where
\begin{equation}
\zeta L=\frac{1}{8\pi}\frac{N_g\lambda_0^2}{n^2 A}\frac{\gamma^2}{\Delta^2}.
\label{etaL}
\end{equation}

\section{Requirements for single-photon sensitivity and low loss; multipass arrangement}

Single photon sensitivity requires $\sqrt{\eta N_p} \phi > 1$, with $\phi$ given by Eq. (\ref{phi}) with an additional factor $1/2$ assuming that the excited state population is transferred to another ground state to minimize loss and noise. Here $\eta$ is the retrieval efficiency of the AFC memory \cite{Afzelius2009},
\begin{equation}
\eta=(1-e^{-d/F})^2 e^{-\frac{\pi^2}{2 \ln 2 F^2}}.
\end{equation}
Assuming $N_p=N_g$ as discussed in section \ref{probestorage}, this puts a lower bound on the number of atoms in $g$, \begin{equation}
N_g > \frac{1}{\eta \phi^2}=\frac{1}{\eta}\left(\frac{8 \pi n^2 A \Delta}{ \lambda_0^2 \gamma}\right)^2,
\label{Ng}
\end{equation}
and hence, using Eq. (\ref{etaL}), a lower bound on the loss experienced by the signal:
\begin{equation}
\zeta L = \frac{ N_g \lambda_0^2 \gamma^2}{8 \pi \eta n^2 A \Delta^2} > \frac{8\pi n^2 A}{ \eta \lambda_0^2}.
\end{equation}
Even for very small cross sections of order $\lambda_0^2/n^2$, this loss is $8 \pi/\eta \gg 1$, which is too high for a nondestructive measurement.

This problem can be overcome by using a multipass arrangement, where $m$ is the number of passes the signal makes through the medium. In this case the phase shift $\phi$ in Eq. (\ref{phi}) and the relation for the loss $\zeta L$ in Eq. (\ref{etaL}) are both multiplied by $m$. However, as a consequence of the former, the lower bound on $N_g$ of Eq. (\ref{Ng}) is multiplied by $\frac{1}{m^2}$, which finally leads to a modified bound on the total loss,
\begin{equation}
\zeta L > \frac{8 \pi n^2 A}{m \eta \lambda_0^2},
\label{loss}
\end{equation}
which can be much less than one for sufficiently many passes.  Requiring small signal loss $\zeta L \lesssim 0.1$,  Eq. (\ref{loss}) gives a condition on $m$,
\begin{equation}
m > 80 \pi/\eta,
\label{cond1}
\end{equation}
where we have assumed a small waveguide, $A=\lambda_0^2/n^2$. Implementing $m \gg 1$ in practice requires low-loss switches. However, an analogous effect can also be achieved by using a cavity. The main difference is that a cavity enhances the signal field rather than the interaction time, which reduces the requirements on the storage time for the probe compared to a multi-pass scenario. Here we focus on the multi-pass case for simplicity.

Another condition on $m$ follows from Eq. (\ref{Ng}), which in the multi-pass case can be rewritten as
\begin{equation}
d > \frac{128 \pi^2 \Delta^2}{n_t \gamma^2 m^2 \eta},
\label{d}
\end{equation}
where we have introduced the optical depth $d$, which for a small waveguide as above, and for an AFC where each tooth corresponds to one radiatively broadenend line, is related to the total number of atoms $N$ as $N=n_t d$, where $n_t$ is the number of teeth in the comb. In deriving Eq. (\ref{d}) we have again assumed $N_g=N/2$. Eq. (\ref{d}) yields a condition on the number of passes,
\begin{equation}
m > \frac{8 \sqrt{2} \pi \Delta}{\sqrt{n_t \eta d} \gamma}.
\label{npDelta}
\end{equation}
We now rewrite the detuning $\Delta=f n_t F \gamma$, where $F=\Delta_m/\gamma$ is the finesse of the AFC and $f$ is a factor greater than one that assures that the signal is sufficiently far detuned from the AFC (whose total width is $n_t F \gamma$). This yields
\begin{equation}
m > 8 \sqrt{2} \pi f F \sqrt{\frac{n_t}{d \eta}}.
\label{cond2}
\end{equation}

Eq. (\ref{npDelta}) also yields a condition on the number of passes as a function of the desired signal bandwidth $B$. For $B$ expressed in Hz one has $B=\Delta/(2 \pi f)$, where the factor $f$ again ensures that the signal is off-resonant. This gives
\begin{equation}
m > \frac{16 \sqrt{2} \pi^2 f B}{\sqrt{n_t \eta d} \gamma}
\end{equation}

For our material system (Tm ions in Lithium Niobate, which have $\gamma$ of order 9 kHz) all the above conditions are satisfied, for example, \textbf{by} setting $f=3$, $d=30$ as achieved in Ref. \cite{Hedges2010}, $B=500$ kHz, $F=3.2$, $n_t=110$ and $m=930$. The latter is probably impossible for multiple passes using switches, but corresponds to only a moderate-finesse cavity \cite{Faraon2015}. For smaller bandwidth and higher optical depth smaller values of $m$ are sufficient, but note that Eq. (\ref{cond1}) implies $m > 80 \pi$ under all circumstances. The bandwidth could also be increased by working with a system that has a larger $\gamma$, such as Ce ions in appropriate crystals \cite{Wrachtrup} or color centers in diamond.

\end{document}